\RequirePackage{tikz}

\def\reviewing{0}

\if\reviewing1
\documentclass[sn-mathphys]{sn-jnl}
\else
\documentclass[sn-mathphys,iicol]{sn-jnl}		
\fi

\usepackage{amssymb}
\usepackage{amsfonts}
\usepackage{graphicx}
\usepackage{gensymb}
\usepackage{subcaption}
\usepackage{epstopdf, epsfig}
\usepackage{chngcntr}
\usepackage{placeins}
\usepackage{comment}
\usepackage{soul}
\usepackage{color}
\usepackage{siunitx}

\usepackage{tabularx}
\usepackage{array}

\usepackage{enumerate}

\usepackage{url}

\usepackage{setspace}

\def\toUseTikz{1}
\def\highlighting{0}

\if\toUseTikz1
\usepackage{tikz}
\usetikzlibrary{external}
\usepackage{pgfplots}
\usepackage{tikzscale}
\usetikzlibrary{calc}
\usetikzlibrary{positioning}
\usetikzlibrary{patterns}
\usetikzlibrary{decorations.markings}
\usetikzlibrary{plotmarks}
\usetikzlibrary{shapes.misc,shapes.symbols}

\tikzset{external/system call= {pdflatex \tikzexternalcheckshellescape
		-halt-on-error 
		-interaction=batchmode
		-jobname "\image" "\texsource"
		&& pdftops -eps "\image".pdf}}

\pgfplotsset{every axis/.append style={
		label style={font=\footnotesize},
		tick label style={font=\footnotesize},
		legend style={font=\footnotesize}
}}

\pgfplotsset{compat=1.17}
\fi

\if\highlighting1

\fi

\jyear{2022}%

\theoremstyle{thmstyleone}%
\theoremstyle{thmstyletwo}%

\theoremstyle{thmstylethree}%

\begin{document}

\title[Article Title]{Controlled accelerations for Rayleigh--Taylor instability}


\author[1]{\fnm{J. T.} \sur{Horne-Jones}}
\author[1]{\fnm{D. J.} \sur{Glinnan}}
\author*[1]{\fnm{A. G. W.} 
\sur{Lawrie}}\email{Andrew.Lawrie@bristol.ac.uk}
\author[1,2]{\fnm{R. J. R.} \sur{Williams}}

\affil[1]{\orgdiv{Hele-Shaw Laboratory}, \orgname{University of Bristol}, \orgaddress{\street{University Walk}, \city{Bristol}, \postcode{BS8 1TR}, \country{UK}}}

\affil[2]{\orgdiv{Computational Physics Group}, \orgname{Atomic Weapons Establishment}, \orgaddress{\street{Aldermaston}, \city{Reading}, \postcode{RG7 4PR}, \country{UK}}}

\if\reviewing1
\doublespacing
\fi


\abstract{The dynamics of turbulent mixing induced by Rayleigh--Taylor instability are heavily dependent on the acceleration experienced by the fluids and the frequency content of the initial interface between them. Both are readily controllable in numerical simulations, but in experimental studies are difficult to influence and adequately diagnose. In this paper we present the CAMPI apparatus, an experimental facility for study of low Atwood number Rayleigh--Taylor instability with highly controllable, complex acceleration histories. The apparatus provides unique and novel capability for the experimental study of variable acceleration Rayleigh--Taylor instability with fully miscible fluids and at a scale suitable for high resolution optical diagnostics. We present experimental results of initially single mode instability evolution through two stepwise acceleration reversals, a case termed Accel-Decel-Accel, demonstrating the ability of the apparatus to accurately generate a prescribed acceleration history. We observe the behaviour predicted by previous numerical studies, with instability growth reaching a terminal velocity in the first episode of acceleration, followed by a shrinking and homogenisation of the mixing region throughout deceleration, and unstable growth from a multi-frequency initial condition during the second acceleration. We present the CAMPI apparatus to the  field as a much needed resource of ground truth data on the behaviour of Rayleigh--Taylor instability across a broad range of regimes.}


\maketitle

\section{Introduction}

\subsection{Classic Rayleigh--Taylor instability}

Rayleigh--Taylor instability is a flow that develops from a horizontal interface between two fluids of differing density: a dense fluid positioned above the interface and a fluid with lower relative density below \cite{rayleigh1900scientific,taylor1950instability}. Initial work \cite{taylor1950instability} considered the behaviour of early-time growth, but the majority of subsequent research has focused on the non-linear self-similar regime that follows, as first investigated experimentally by Read \cite{read1984RTexperimental} and numerically by Youngs \cite{youngs1984RTnumerical}. This form of self-similar development, under constant applied acceleration $\hat{g}$, takes the form of a quadratic in time,
\begin{equation} \label{eq:selfSimilar}
h = \alpha \hat{g} A_{\rm t} t^2,
\end{equation}
where $h$ is the height of a region around the original interface containing a mixture of fluid from above and below the interface, $\alpha$ is a scaling constant, and $t$ is time. The Atwood number, $A_{\rm t}$, is a non-dimensional measure of density difference defined as
\begin{equation}
A_{\rm t} = \frac{\rho_{\rm h} - \rho_{\rm l}}{\rho_{\rm h} + \rho_{\rm l}},
\end{equation}
where $\rho_{\rm h}$ is the density of the more dense fluid and $\rho_{\rm l}$ is the density of the less dense fluid. The classical Rayleigh--Taylor instability has been the subject of extensive investigation; details are available in comprehensive review articles \cite{schilling2020progress,boffetta2017incompressible,zhou2017rayleighI,zhou2017rayleighII}. 

\subsection{Initial conditions in Rayleigh--Taylor}

It is well documented \cite{aslangil2016numerical,dimonte2004dependence,horne2020aspect,olson2009experimental,ramaprabhu2004initialization,ramaprabhu2005numerical} that initial conditions play a significant role in the development of Rayleigh--Taylor instability, despite the emergence of self-similar behaviour at later times. It has been known since the work of Taylor \cite{taylor1950instability} that growth-rates of individual modes are sensitive to the modal wavenumber, with the overall development comprised of the superposition of all modes present in the system. Youngs \cite{youngs1984RTnumerical} initialised his simulations with small random perturbations to achieve good agreement with the experiments of Read \cite{read1984RTexperimental}, approaching the self-similar growth predicted by (\ref{eq:selfSimilar}). Characterisation of initial conditions has proven to be an essential part of understanding the initial transient response before asymptotically self-similar behaviour emerges. A further motivation for new study is the pursuit of carefully validated ground-truth experimental data to compare with numerical data sets \cite{dalziel1999self,mueschke2009investigation,ramaprabhu2004initialization,olson2009experimental}. There is a long history in the field of trajectory divergence between experiment and notionally equivalent simulations of Rayleigh--Taylor \cite{youngs2017rayleigh}, because experimental initial conditions have proven especially difficult to characterise \cite{banerjee20093d}. Choice of experimental configuration is the dominant influence on initial conditions and their control. We sought to improve on prior work by enabling more precise control of initial conditions and enhancing the generality of subsequent acceleration time-histories. Detailed overviews on experimental configuration can be found in recent review articles \cite{zhou2017rayleighI,banerjee2020rayleigh}.

The most popular experimental methods can be grouped into four categories: (1) the removal of a barrier supporting the denser fluid against the acceleration field (usually gravity); (2) a channel whose flow is split by a plate upstream, but allowed to mix downstream; (3) a statically stable density stratification that is overturned to become unstable; and (4) a change to the acceleration imposed on a statically stable density stratification such that is becomes unstable. Removal of thin, solid, barriers \cite{linden1994molecular,wykes2014efficient} causes shear at the interface due to shedding of the boundary layer, and composite barriers designed to avoid shear \cite{dalziel1993rayleigh,dalziel1999self,lawrie2011rayleigh} have non-negligible thickness that induces a significant and long-term mean flow. When using flow channel methods \cite{snider1994rayleigh,banerjee2006statistically,banerjee2010detailed}, the boundary layers between fluids and plate are advected into the mixing region, precisely co-located with the interface that is the primary object of study, and this pollutes important features of the initial condition. Overturning an initially stable stratification \cite{andrews1990RTexperimental2D,horne2020aspect} is limited to geometrically thin systems because the fluid inertia induces a relative rotation of the interface, which if it were to persist for long enough would develop into Kelvin--Helmholtz instability in the `thin' direction, the dynamics of which then modify the initial condition for the evolution of subsequent Rayleigh--Taylor instability. Imposing an acceleration \cite{read1984RTexperimental,waddell2001experimental,wilkinson2007experimental,olson2009experimental} is the only approach where Rayleigh--Taylor instability is free to grow from an initially quiescent body of fluid. While the instability growth is limited by the distance over which a suitable acceleration can be sustained, all significant work done on the fluid is achieved by imposed body forces, which can be precisely measured even if never fully controlled. This offers significant advantages when seeking to validate numerical predictions. Furthermore, carefully performed oscillatory body forcing \cite{read1984RTexperimental,waddell2001experimental,wilkinson2007experimental,olson2009experimental} can establish well-bounded interfacial perturbations that simplify the task of cross-validation.

\subsection{Variable acceleration in Rayleigh--Taylor instability}

An important degree of freedom offered by imposed acceleration is that the mean body force may be varied during the evolution of Rayleigh--Taylor instability, and this is representative of some industrial applications \cite{kilkenny1994review,betti1998growth,remington2019rayleigh} and major astrophysical events \cite{cabot2006reynolds}. The majority of academic research on Rayleigh--Taylor only considers cases with constant acceleration. Our experimental configuration is the first to provide ground-truth data for variable acceleration cases across a broad operating regime.

The most noteworthy prior experiments are those of Dimonte and Schneider \cite{dimonte1996turbulent} and Dimonte et al. \cite{dimonte2007rayleigh}, both of which used a linear electric motor (LEM) to apply the unstable acceleration. The latter study also incorporated mechanical leaf springs to provide stable deceleration before a second period of unstable acceleration driven by the LEM; an Accel-Decel-Accel acceleration profile. The experiments showed that the dominant structures that grow in the initial unstable phase then shred themselves during the stable deceleration phase, and growth during the second unstable phase is slower due to the well mixed state of the mixing region, which may be considered as a non-quiescent initial condition for the subsequent acceleration. However, the capability of the apparatus restricted these earlier studies to use of immiscible fluids with relatively high Atwood numbers, a small test cell volume, and accelerations of short duration. These restrictions in scale augmented the behavioural complexities of surface tension and molecular diffusion and limited the evolution of the instability in both phases of unstable acceleration.

The behaviour observed during the episode of stable deceleration in the above experiments is corroborated by comparable numerical studies \cite{ramaprabhu2013rayleigh,aslangil2016numerical,aslangil2022rayleigh}. These studies also report eventual convergence to self-similar growth during the second episode of unstable acceleration. Aslangil et al. \cite{aslangil2022rayleigh} conclude that the response of Rayleigh--Taylor instability during this episode depends on the behaviour during deceleration and the timing of the second acceleration reversal with respect to the phase angle of any oscillatory dynamics. Consequently, prediction of the response in the second episode of acceleration requires characterisation of the internal wave dynamics present during deceleration.

\subsection{Organisation}

In this paper we investigate low Atwood number Rayleigh--Taylor instability under Accel-Decel-Accel acceleration profiles, initialised with prescribed single mode initial conditions. These experiments are conducted using the CAMPI (\textbf{C}ontrolled \textbf{A}cceleration for \textbf{M}ulti-\textbf{P}hase \textbf{I}nstabilities) apparatus, which was designed to facilitate experimental investigation of Rayleigh--Taylor instability across a broad range of acceleration time-histories, and is introduced in this paper. The apparatus facilitates the control of initial conditions and the imposition of a specified acceleration history on a stratified fluid system with Atwood number up to $A_{\rm t} = 0.27$, diagnosed by high speed, well-resolved optical diagnostics. We describe our diagnostic post-processing techniques, then present our observations of fluid mixing behaviour during episodes of deceleration and subsequent re-acceleration some time after instability had been allowed to develop.

The paper is organised as follows: section \ref{Rig} presents the CAMPI variable acceleration apparatus, section \ref{ExpMethod} outlines the experimental and post-processing methods used, section \ref{Results} presents results and analysis from the experiments, and conclusions are drawn in section \ref{Conclusions}.

\section{The CAMPI apparatus} \label{Rig}

The CAMPI variable acceleration apparatus was designed to facilitate well-diagnosed experimental study of Rayleigh--Taylor instability across a broad operating regime. It uses a closed loop electrical winch to raise and lower a fluid-filled cuboidal tank along a $12\mathrm{\,m}$ vertical track. In the experiments presented here, the tank contains a statically stable horizontal density interface. The winch is capable of peak accelerations at the density interface of $5$ times the force of terrestrial gravity, $g$, with $350\mathrm{\,Nm}$ of torque delivery from the electric motor over the full speed range. An optical diagnostic system is mounted in the  moving reference frame to concurrently excite a pair of onboard lasers, record an image sequence, and record acceleration. This section describes the motivation for the design choices made in developing the apparatus. Section \ref{DesignMotivation} discusses the system design requirements, then the design of the accelerator, the onboard diagnostic systems, and the control systems are described in sections \ref{RigAccelerator}-\ref{RigSoftware} respectively. The reader is referred to Horne \cite{horne2022rayleigh} (chapter 3) for further details on all aspects of the design of the CAMPI apparatus.

\subsection{Design motivation} \label{DesignMotivation}

In the non-linear regime of Rayleigh--Taylor instability, growth of the mixing layer takes the form $h(t)=\alpha A_{\rm t} \hat{g} t^2$, where $\hat{g}$ is the constant applied acceleration experienced by the fluid system, and development of the mixing layer is therefore sensitive to the magnitude and duration of this acceleration. The primary limitation in using acceleration to initiate Rayleigh--Taylor instability is the duration for which continuous acceleration can be achieved, since the distance over which the system can be moved is limited by the space available. Interfacial development during unstable acceleration must be well-resolved at the diagnostic resolution to generate results with sufficient signal to noise ratio for meaningful scientific analysis. The relationship between the achievable acceleration and the instability growth therefore provides a key design constraint.

\begin{figure}
	\centering
	\if\toUseTikz1
	
	\tikzsetnextfilename{GrowthVsAparGrav}
	\begin{tikzpicture}
\begin{axis}[
    name=plot1,
    axis on top,
    width=1.0\linewidth,
    xmin=0, xmax=100, ymin=0, ymax=0.9,
    x label style={at={(axis description cs:0.5,-0.05)},anchor=north},
    y label style={at={(axis description cs:-0.08,0.5)},anchor=south},
    xlabel={$\hat{g}_{max}({\rm\,m\,s^{-2}})$},
    ylabel={$\frac{h}{\alpha A_t s_{total}}$},
    tick label style={font=\normalsize},
    label style={font=\large},
    clip=false, 
]

    \addplot[thick,blue] graphics[xmin=0,ymin=0,xmax=100,ymax=0.9] {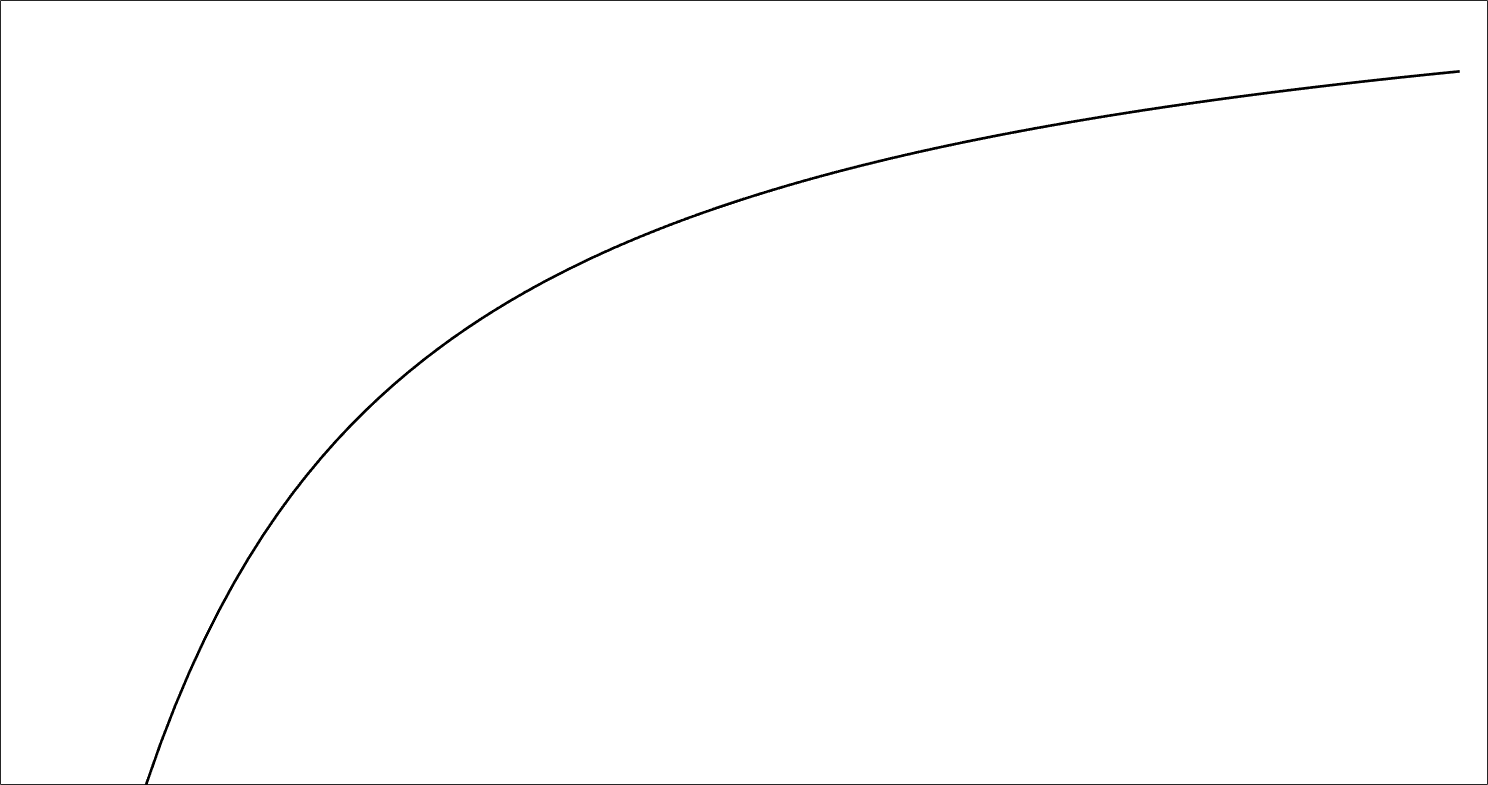};

    \addplot[only marks,mark=*] coordinates {(9.8,0)};       \coordinate (gpt) at (axis cs:9.8,0);
    \addplot[only marks,mark=*] coordinates {(49,0.665)};    \coordinate (gmaxpt) at (axis cs:49,0.665);
    \addplot[only marks,mark=*] coordinates {(49,0)};        \coordinate (gbase) at (axis cs:49,0);
    \coordinate (gmaxpt2) at (axis cs:60,0.70);

    \draw[dotted] (axis cs:49,0.665) -- (axis cs:49,0);
\end{axis}

\node[below,yshift=-1mm] at (gpt) {$g$};            
\node[below,yshift=-1mm] at (gmaxpt2) {$\hat{g}_{\textrm{max}}^{\textrm{(design)}}$}; 

\end{tikzpicture}

	\else
	\includegraphics[width=\linewidth]{GrowthVsAparGrav.eps}
	\fi
	\caption{The variation of Rayleigh--Taylor growth with apparent gravity during a maximal simple acceleration profile spanning the full track length.}
	\label{fig:heightVsApparentGravity}
\end{figure}

We consider a central design case where a constant maximal force downwards is followed by the same magnitude force upwards, such that the tank translates along the full length of the track. The accelerations of the tank during each stage of motion are $a_{\rm{down}} = \hat{g}_{\rm{max}} + g$ and $a_{\rm{up}} =  -\hat{g}_{\rm{max}} + g$, where $\hat{g}_{max}$ is the component of acceleration resulting from the applied force. We determine the relationship between $\hat{g}_{\rm{max}}$ and the duration of downward acceleration for a tank translation, $s_{\rm{total}}$, and obtain the prediction for the resultant Rayleigh--Taylor growth
\begin{equation} \label{eq:growthVsGHat}
h=\alpha A_{\rm t} s_{\text{total}} \frac{\hat{g}_{\text{max}}-g}{\hat{g}_{\text{max}}+g}.
\end{equation}
This relationship is shown graphically in figure \ref{fig:heightVsApparentGravity}, from which the diminishing returns from greater applied accelerations can be seen. Figure \ref{fig:heightVsApparentGravity} was used to select a design maximum applied acceleration of $\hat{g}_{\text{max}} = 5g = 49{\rm\,m\,s^{-2}}$.

The experimental tank dimensions are a compromise to limit moving mass while retaining high diagnostic quality for key design cases. Our experimental approach intends to study interfacial development in some approximation to free space in the lateral direction, but for this to be possible over the full height of the tank would require a tank width significantly exceeding the instability height. Practical considerations lead us to select a 2:1:1 height to width to depth ratio with internal dimensions $400 \times 200 \times 200 \mathrm{\,mm}$. The design mass of the whole moving assembly is set at $45 \mathrm{\,kg}$. To achieve the maximal acceleration described above, the propulsion system applies a net force, $F_{\text{load}} = 2207 \mathrm{\,N}$, on the moving assembly over a speed range that peaks at $v_{\text{max}} = 22.2 {\rm\,m\,s^{-2}}$.

\subsection{Accelerator} \label{RigAccelerator}

We sought precise bi-directional control to enable consecutive episodes of destabilising and stabilising constant accelerations, with trivial generalisation to smoothly varying single and multi-mode oscillatory acceleration histories. The majority of previous experimental work on Rayleigh--Taylor by acceleration of the system produced single episodes of approximately constant acceleration using, variously, stretched rubber tubing \cite{emmons1960taylor}, compressed air \cite{cole1973experimental}, rocket motors \cite{read1984RTexperimental}, and falling weights \cite{waddell2001experimental}. Dimonte et al. \cite{dimonte2007rayleigh} generated multiple segments of acceleration in a longer time-history by using a combination of a linear electric motor and a mechanical leaf spring system. Our solution is a closed loop winch system in which the winch rope, a synthetic fibre weave chosen for its tensile stiffness, simultaneously winds on and off a helically grooved aluminium drum on high-speed bearings, and the moving carriage is located in series with the winch loop. A large servomotor and motor inverter drive combine to deliver a torque of $350 \mathrm{\,Nm}$ to the drum for up to $5 \mathrm{\,s}$ with rotational speed peaking at $2000 \mathrm{\,rpm}$. The motor and drum are connected by a torsionally stiff but laterally flexible coupling, resulting in a low backlash drive system that improves the accuracy and controllability of our accelerations. An industrial fail-safe brake is mounted directly onto the drum shaft.

The track is mounted onto a $13 \mathrm{\,m}$ tall universal column. It has a twin C-shaped profile, with six running faces, and provides a vertical guide to which the carriage is laterally and rotationally constrained but along which it can freely run. The carriage wheels, wheels originally designed for speed-skating, have a small amount of compliance that permits an interference fit between the running gear and track that helps to reduce vibration.

\begin{figure*}
	\centering
	\if\toUseTikz1
	
	\tikzsetnextfilename{FullRigDrawing}
	\input{FullRigDrawing.tikz}

	\else
	\includegraphics[width=\linewidth]{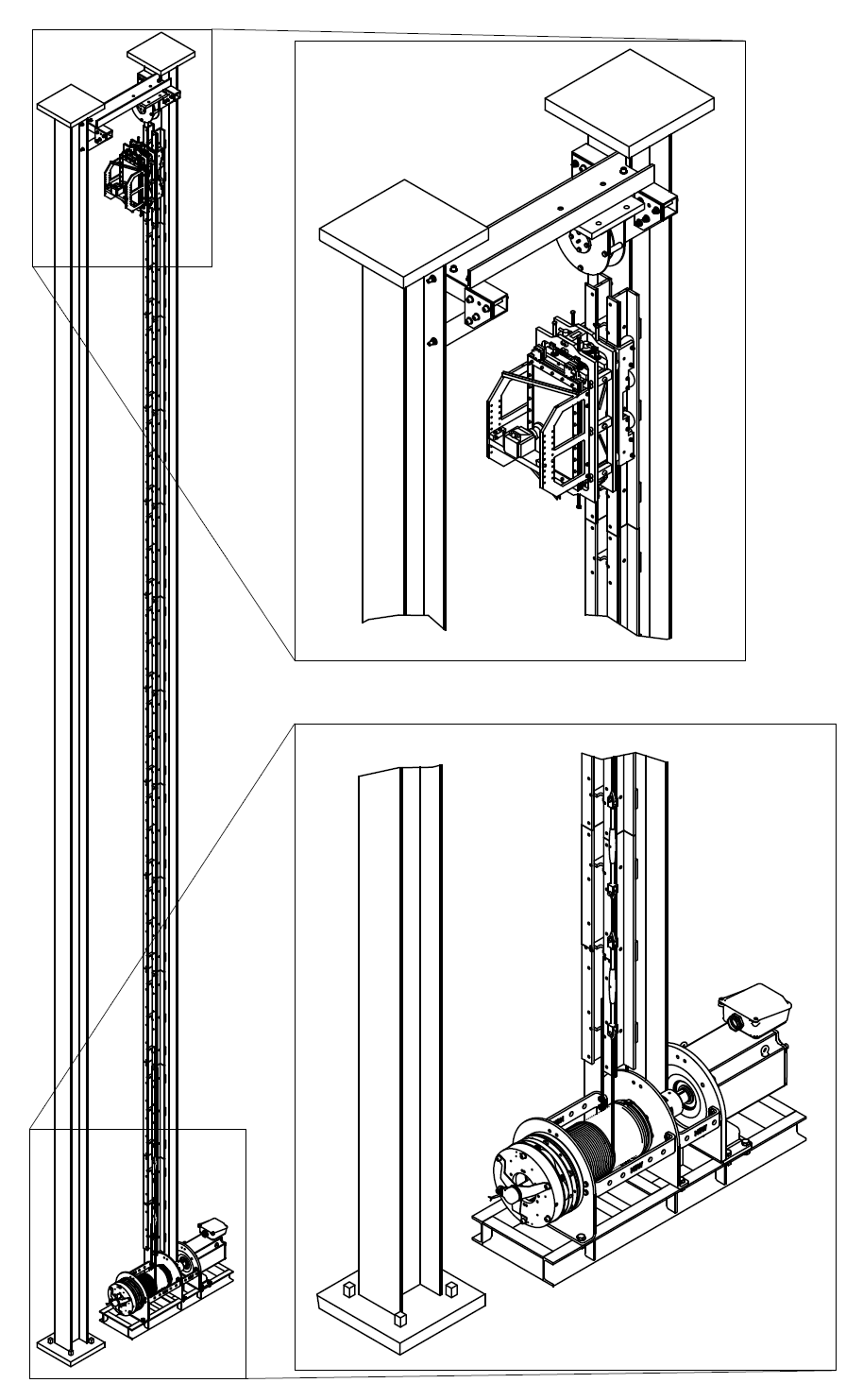}
	\fi
	\caption{The CAMPI variable acceleration apparatus. The propulsion is provided by a closed loop winch system, with drum, motor, and brake located on a single shaft at the base of the apparatus, and return pulley located at the top. The apparatus is mounted on a pair of universal columns that span the full height of the Hele-Shaw Laboratory. The experimental setup is contained within a shuttle assembly that runs along a track spanning the full height of the apparatus; $12.6 \mathrm{\; m}$ between the two rotational axes.}
	\label{fig:FullRigDrawing}
\end{figure*}

The rope is given a static tension of $3000 \mathrm{\,N}$, greater than the maximum accelerative loading, to ensure that there are no design conditions in which the rope could become slack. A CAD view of the whole apparatus is shown in figure \ref{fig:FullRigDrawing}.

\subsection{Moving reference frame diagnostic systems} \label{RigDiagnostics}

\begin{figure}
	\centering
	\if\toUseTikz1
	
	\tikzsetnextfilename{DiagnosticSystems}
	\input{DiagnosticSystems.tikz}

	\else
	\includegraphics[width=\linewidth]{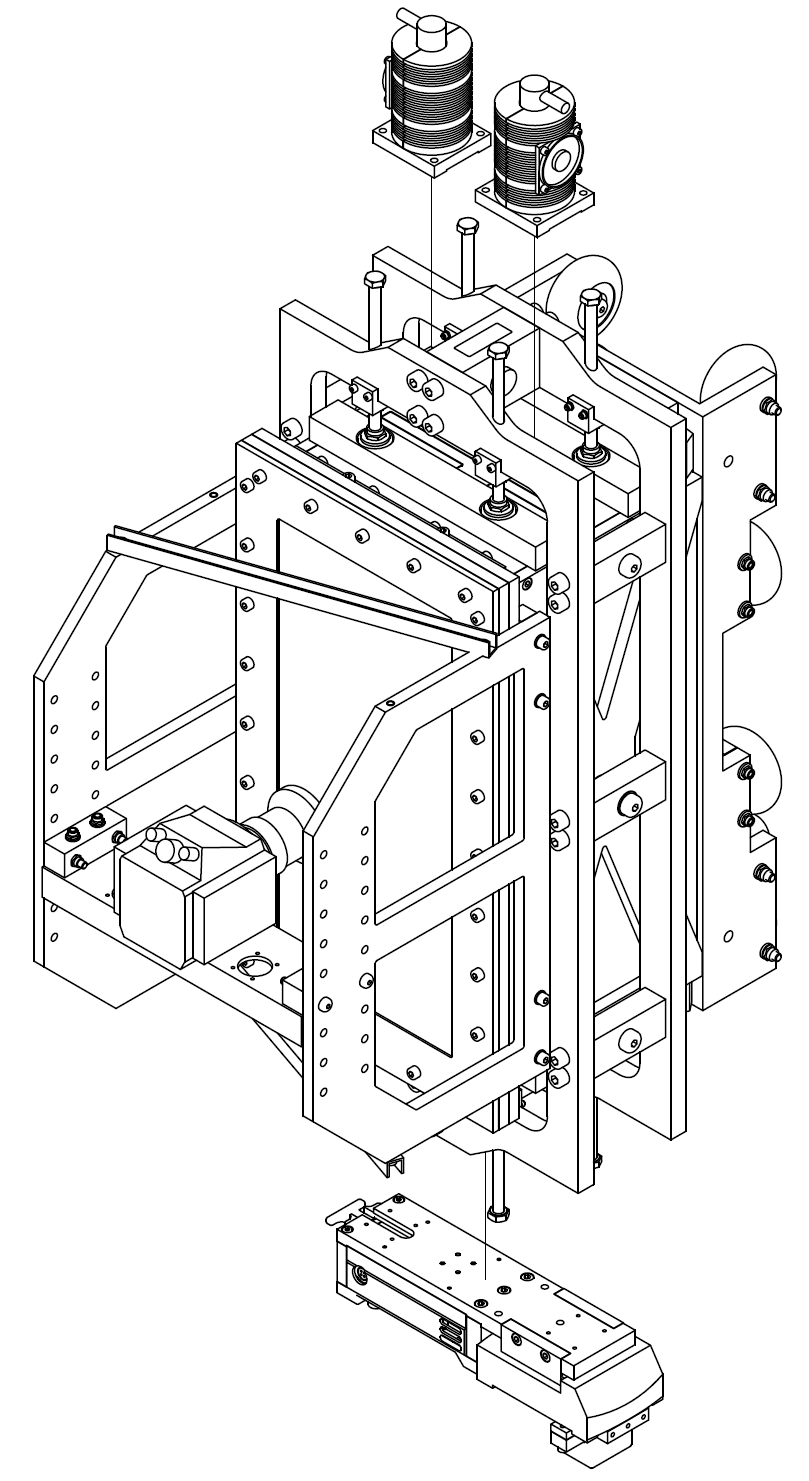}
	\fi
	\caption{The moving reference frame onboard diagnostic systems.}
	\label{fig:DiagnosticSystems}
\end{figure}

Acceleration-based Rayleigh--Taylor experiments require synchronous diagnostics measuring system input, the acceleration history, and the output, the fluid response. The design of the CAMPI apparatus requires that all diagnostic systems are in the moving reference frame; mounted on the tank assembly. The various constituent parts of the systems are shown in figure \ref{fig:DiagnosticSystems}.

The acceleration history is measured using a digital accelerometer chip. This measures over a $\pm 8g$ range at sample rates of up to  $3200 \mathrm{\,Hz}$, and is mounted with a short and direct contact path to the base of the tank to minimise signal noise due to relative vibration.

We use Planar Laser Induced Fluorescence (PLIF), to measure concentration of the active scalar. The PLIF diagnostic uses $520 \mathrm{\,nm}$, $300 \mathrm{\,mW}$ diode laser modules to provide mid-plane illumination of the experiment. The modules are fitted with optics to generate $90 \degree$ fan angle single line output with plane thickness varying between $0.5 - 2.0 \mathrm{\,mm}$ over the height of the tank. The wavelength is selected for use with Rhodamine 6G fluorescing dye, which has peak absorption at $525 \mathrm{\,nm}$ and peak emission at $554 \mathrm{\,nm}$ \cite{wurth2012determination}. A low-pass optical filter, with cut-off wavelength of $530 \mathrm{\,nm}$, is mounted outside of the Perspex front face of the tank and is used to filter out the majority of the laser light, while transmitting the light emitted by the dye. Experiments on the CAMPI apparatus have an active duration of $\mathcal{O}(1 \mathrm{\,s})$ and therefore use a high speed camera capable of $4000 \mathrm{\,fps}$ capture of 12-bit depth colour images at $1440 \times 1024$ pixel resolution. The camera is installed in front of the Perspex front face of the tank, requiring use of a fisheye lens to provide full field of vision of the illuminated plane. The diagnostic output, after post-processing, is of images of volume fraction at the illuminated plane.

A Beaglebone AI single-board computer provides all onboard compute capability. Its Texas Instruments AM5729 chip incorporates a dual $1.5 \mathrm{\,GHz}$ Arm Cortex A15 CPU and two dual-core 32-bit $200 \mathrm{\,MHz}$ Programmable Real-Time Unit and Industrial Communication SubSystems (PRU-ICSS). The CPU provides conventional compute facilities and the PRUs provide microcontroller architectures with no risk of operating system interrupts. The Beaglebone includes a pair of 46-pin headers that incorporate general-purpose input/output (GPIO) pins, a specific set of which are also available from each PRU with single cycle ($5 \mathrm{\,ns}$) read/write access.

A bespoke two-sided PCB is used to house all power and component interface circuitry and facilitates very low latency binary signal communication between PRU cores. The onboard systems are powered by a Lithium-ion battery pack that supplies this PCB.

\subsection{Control systems} \label{RigSoftware}

Three pieces of interconnected hardware control the apparatus: a PLC for the motor inverter drive; a Beaglebone AI single board computer for the onboard diagnostics; and a desktop computer providing the human-machine interface. Specific control software runs of each piece of hardware and the three control systems communicate with each other over a private network using the MODBUS TCP/IP client/server protocol. The motor controller and diagnostic control systems are configured as two separate servers and the user-machine interface is a client to both. Control logic is based on 32-bit `control words' and `status words' that are passed over the MODBUS connection.

The control systems are constructed with varying levels of real-time determinism. The user interface and the MODBUS server functionality run in untimed loops aimed at minimising reaction time to inputs without requiring strict temporal determinism. For the remainder of the control logic, each control system runs its own timing thread that coordinates other threads executing various control functions, all of which are kept sufficiently lean to reliably execute within the timing deadlines.

The primary function of the motor control system is to provide reference velocities, integrated from a user-defined valid acceleration profile, as input to a proportional–integral–derivative (PID) controller that sets drive output. The control logic also incorporates a number of safety features, including hardware and software track limit switches, motor current limiting, position control loop error monitoring, and position-dependent velocity limits, automatically overriding preceding operation if required.

The onboard control systems are split between the CPU and the four PRU cores of the Beaglebone. The CPU runs the network server for the onboard diagnostics and manages the behaviour of the PRUs. Runtime communication between the CPU and PRU software is facilitated by the Remote Processor Messaging (RPMsg) framework. The hard real-time diagnostic functionality is distributed across the four PRU cores. The first core outputs a precise $2 \mathrm{\,MHz}$ clock signal onto a single bit bus that is parsed as input to each of the other cores. The second core, the scheduler, is responsible for generation of timing signals for laser and camera control and for accelerometer sampling. The scheduler is provided with timing signal definitions by the CPU software and counts clock signal periods in order to generate the timing signals. The laser and camera PRU core takes a coarse and a fine resolution timing signal as input from the scheduler, and uses these to construct modulation signals for the lasers and to trigger frame capture on the camera. This approach offers considerable flexibility, such as bursts of frame triggering and sequenced modulation of several laser modules. The final core manages an SPI interface with the three-axis accelerometer chip, synchronising the data capture with camera frame capture.

The human-machine interface allows the user to issue requests to the motor controller server and to the onboard diagnostics server. The timing structure outlined previously is used to poll both servers at frequent, regular intervals to forward user requests and obtain live system status. The MODBUS clients and user interface are all executed in dedicated threads, interfacing via shared memory to communicate user requests and statuses between all threads. The user interface provides the facility for construction and validation of acceleration profiles, details of which can subsequently be transferred to the two servers before apparatus operation.

\subsection{Controller design} \label{Controller design}

\begin{figure} 
	\centering
 \includegraphics[width=\linewidth]{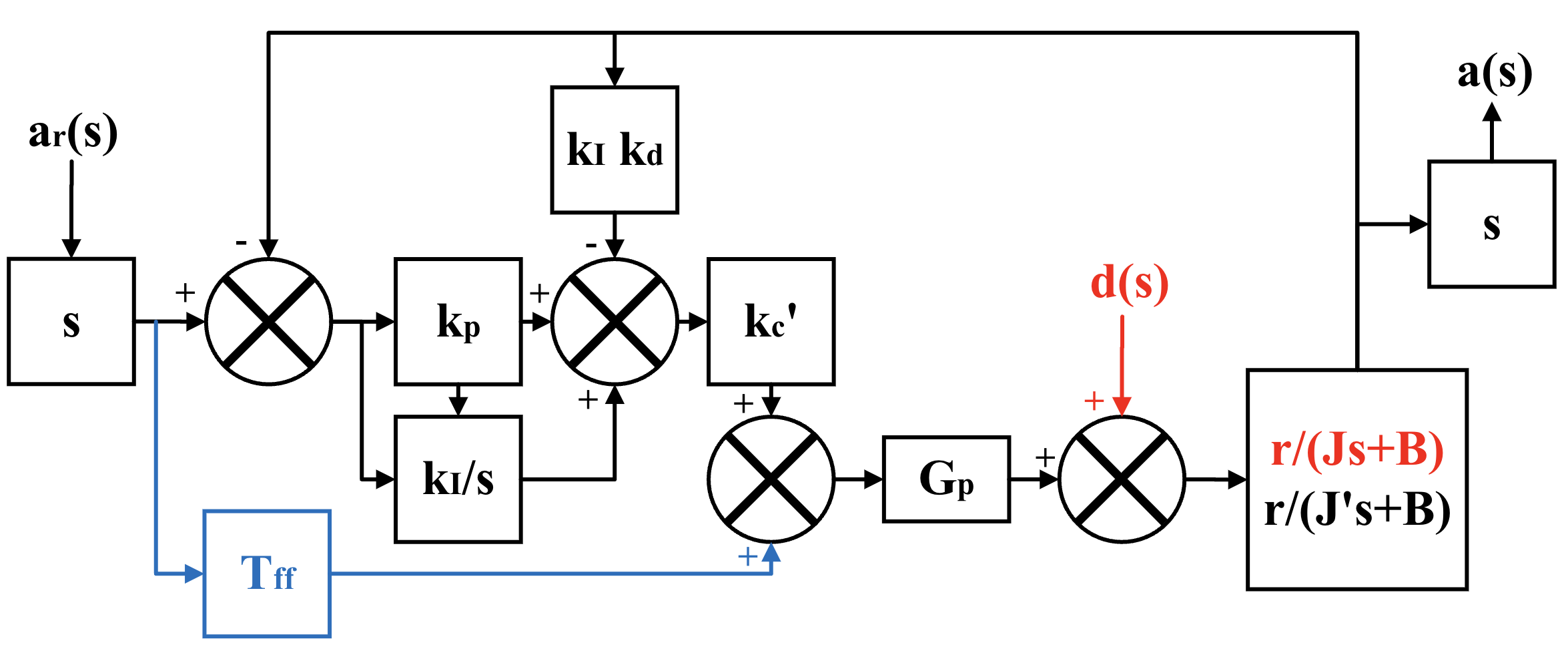}
        \newline
	
	\caption{The closed loop control block diagram showing the improved system (black), the original system (black + red) and the implementation of torque feed forward (blue).}
	\label{fig:Control Diagram}
\end{figure}

\begin{figure*} [t!]
	\centering
	\if\toUseTikz1
	
	\tikzsetnextfilename{AccelData}
	\begin{tikzpicture}

\begin{axis}[
axis on top,
width=\linewidth,
height=0.4\linewidth,
xmin=0,
xmax=10.5,
ymin=-40,
ymax=30,
xlabel={time (s)},
ylabel={$\hat{g}\,(\mathrm{m\,s^{-2}})$},
xtick={0.5,1.5,2.5,3.5,4.5,5.5,6.5,7.5,8.5,9.5,10.5},
    xticklabels={0,1,2,3,4,5,6,7,8,9,10},
    tick label style={font=\normalsize},
    label style={font=\large},
]

\addplot[thick,blue] graphics[xmin=0,ymin=-40,xmax=10.5,ymax=30] {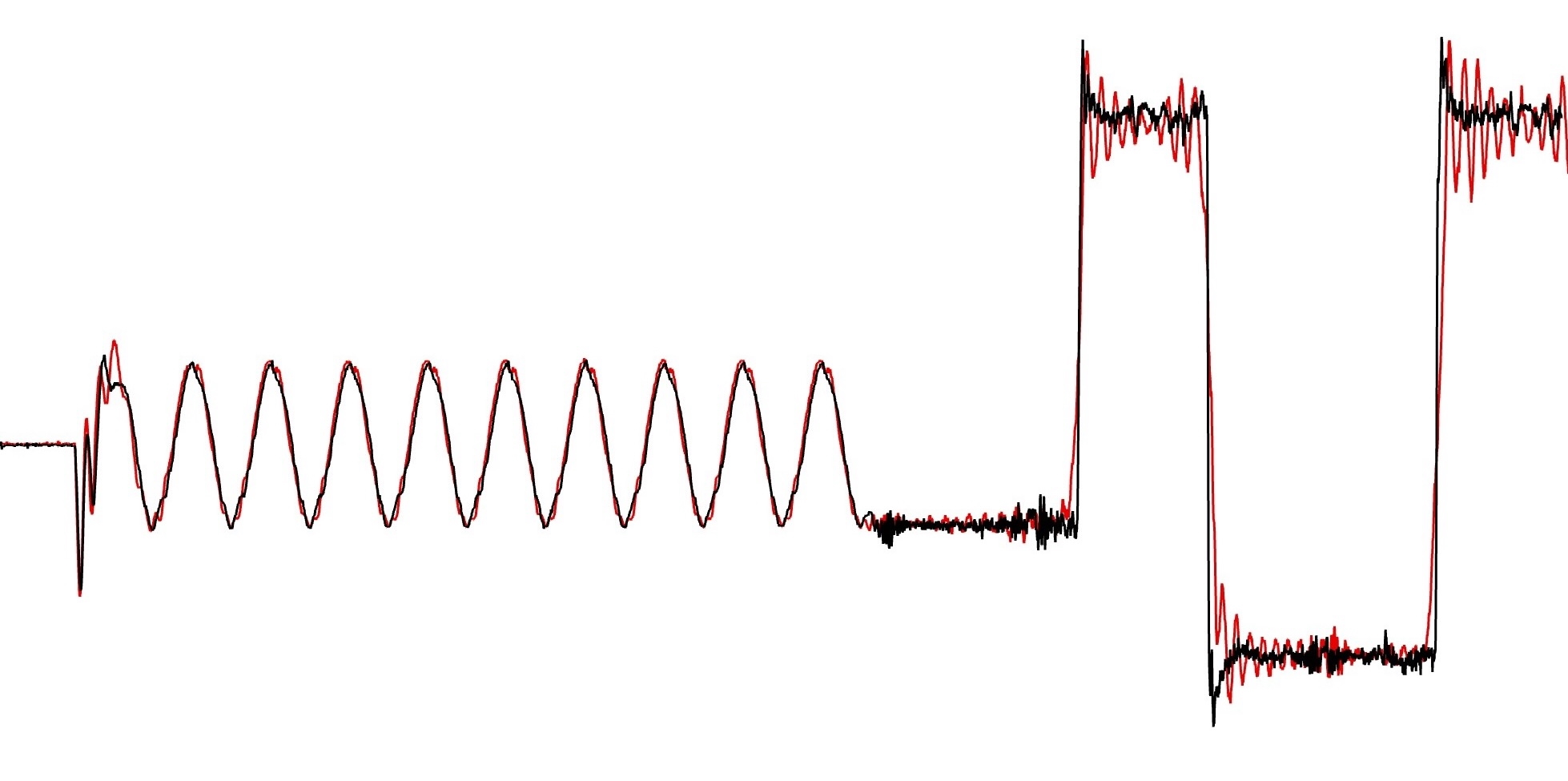};

\draw[color=blue,fill=blue,opacity=0.05] (axis cs:0.5,-40) rectangle (axis cs:5.8,30);
\draw[color=green,fill=green,opacity=0.1] (axis cs:7.27-0.05,-40) rectangle (axis cs:8.145-0.05,30);
\draw[color=yellow,fill=yellow,opacity=0.1] (axis cs:8.145-0.05,-40) rectangle (axis cs:9.70-0.05,30);
\draw[color=green,fill=green,opacity=0.1] (axis cs:9.70-0.05,-40) rectangle (axis cs:10.6,30);

\node[anchor=north] at (axis cs:3.205,-33) {\footnotesize Stable vertical forcing};
\node[anchor=north] at (axis cs:7.685,-33) {\footnotesize Accel};
\node[anchor=north] at (axis cs:8.895,-33) {\footnotesize Decel};
\node[anchor=north] at (axis cs:10.115,-33) {\footnotesize Accel};

\end{axis}

\end{tikzpicture}

	\else
	\includegraphics[width=\linewidth]{AccelData.eps}
	\fi
	\caption{The acceleration time-history as experienced by the fluid before (red) and after (black) the addition of torque feed back and gain scheduling control methods.}
	\label{fig:AccelData}
\end{figure*}

The apparatus operates in a challenging control environment due to three difficulties. Firstly, discontinuities in the desired response give rise to oscillations in the shuttle acceleration. Since the derivative term, $k_d$, acts on the measured angular velocity of the motor rather than the error, an infinite impulse response is avoided \cite{parr1998industrial} (see figure \ref{fig:Control Diagram}). A sharp, finite impulse however, remains in the system and because the mechanical system is low-loss, this gives rise to oscillations. A representative example is shown by the oscillations in the red line plotting the magnitude of acceleration on figure \ref{fig:AccelData}. Secondly, due to variation in the length of the winch rope between the shuttle and the drum and pulley, the natural frequencies of the overall system vary with the shuttle's position, and, to a lesser extent, effects due to friction and motor backlash. 
When the shuttle's velocity changes sign, the system experiences an abrupt transition in the natural frequency of the rope accelerating the tank. In our investigation of an Accel-Decel-Accel profile, the shuttle's dynamics shift from a high natural frequency (short rope length) to a low natural frequency (long rope length).  This temporarily moves the poles of the system's dynamics into the right-half of the complex s-plane, and the system is temporarily unstable.
This manifests in growing oscillations the `decel' episode around $8.4 \mathrm{\,s}$ as shown in figure \ref{fig:AccelData} where the shuttle's velocity changes sign near the bottom of the track.
Thirdly, the gravity term introduces a direction-dependent disturbance into the system, opposing the upward acceleration and assisting the downward acceleration. To address these difficulties we have designed additional control beyond the motor's own default PID.

Torque feed-forward control is implemented to cancel out the dominant accelerative impulse that remains in the system and the gravitational terms, shown by the red components in figure \ref{fig:Control Diagram}, from the system's closed loop control. Since the system is sufficiently rigid, the mass, $m$, of the shuttle induces a moment of inertia $J_{\text{shuttle}}=mR^2$, where \(R\) is the drum's radius. The remaining inertia of the system is denoted $J_{\text{shaft}}$, so we may write the motor torque, $T_{\rm m}$, as 
\begin{equation}
T_{\rm m} = T_{\rm g}+(J_{\text{shuttle}}+J_{\text{shaft}})\frac{d\omega_{\rm n}}{dt}+B\omega_{\rm n},
\end{equation}
where \(T_g\) is the torque due to gravity, \(\omega_n\) is the motor angular velocity and \(B\) is a friction constant. By balancing torques, we have
\begin{equation}
(J_{\text{shuttle}}+J_{\text{shaft}})\frac{d\omega_n}{d t}={maR},
\end{equation}
where $a$ is the acceleration of the shuttle. A feed-forward torque, $T_{\rm ff}$, is given by
\begin{equation}
T_{\rm ff} = -ma_{\rm r}R-T_{\rm g},
\end{equation}
where \(a_{\rm r}\) is a reference signal for the shuttle acceleration that the control system is designed to track. Addition of the feed-forward control pathway changes the motor dynamics to
\begin{equation}
T_{\rm m}' = T_{\rm m} + T_{\rm ff} = mRe_a +B \omega_{\rm n}
\end{equation}
where the acceleration error \(e_a = a - a_{\rm r}\). Expanding the Taylor Series around $a_r$,
\begin{equation}
a = a_{\rm r} + \mathrm{H.O.T.},
\end{equation}
we note that the truncated terms, $\rm H.O.T.$, correspond to $e_a$, which now becomes the sole target of the PID controller. This cancels the acceleration impulse that remains in the system to leading-order and thus significantly improves closed-loop stability. The closed-loop control block diagram is given in figure \ref{fig:Control Diagram}, showing the improved system (black), the original system (black + red) and the implementation of torque feed-forward control. 

\begin{figure}
	\centering
	\if\toUseTikz1
	
	\tikzsetnextfilename{poles}
	\begin{tikzpicture}
\begin{axis}[
    axis on top,
    width=1.0\linewidth,
    xmin=-80,
    xmax=20,
    ymin=-140,
    ymax=140,
    x label style={at={(axis description cs:0.5,-0.05)},anchor=north},
    y label style={at={(axis description cs:-0.08,0.5)},anchor=south},
    xlabel={Real part $(\mathrm{rad\,s^{-1}})$ },
    ylabel={Imaginary part $(\mathrm{rad\,s^{-1}})$ },
    ytick={-140,-70,0,70,140},
    yticklabels={-140,-70,0,70,140},
    xtick={-80,-60,-40,-20,0,20},
    xticklabels={-80,-60,-40,-20,0,20},
    tick label style={font=\normalsize},
    label style={font=\large},
]

    \addplot[thick,blue] graphics[xmin=-80,ymin=-140,xmax=20,ymax=140] {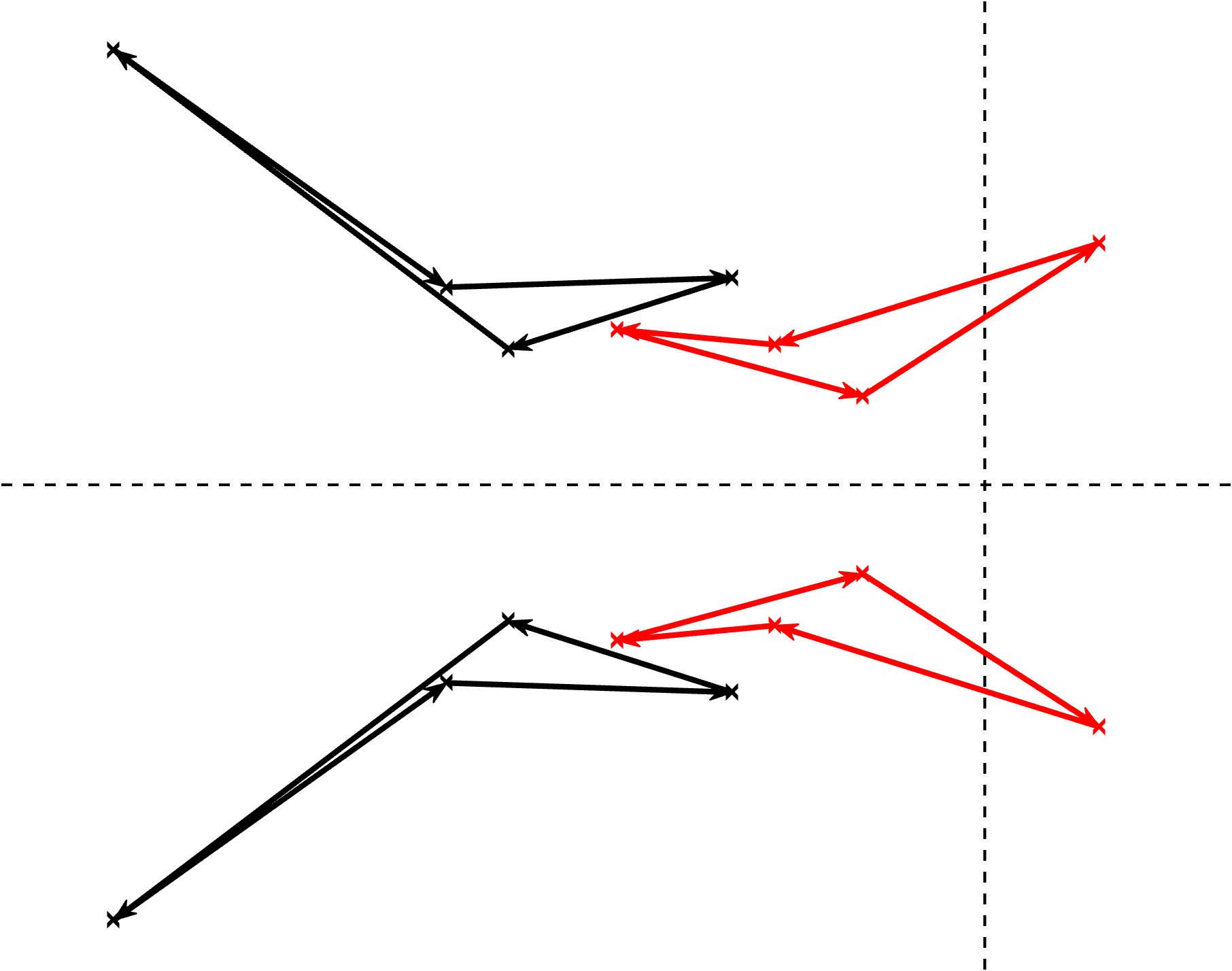};

    \addplot [
        draw=none,
        pattern=north east lines,
        pattern color=gray!50
    ] coordinates {(0,-140) (20,-140) (20,140) (0,140)};

    \addplot[only marks, mark=*, red] coordinates {
        (-9.97, 25.65) (-9.97, -25.65)
        (9.30, 69.81)  (9.30, -69.81)
        (-17.09, 40.54)(-17.09, -40.54)
        (-29.94, 44.88)(-29.94, -44.88)
    };

    \addplot[only marks, mark=*, black] coordinates {
        (-20.55, 59.84)(-20.55, -59.84)
        (-38.76, 39.27)(-38.76, -39.27)
        (-70.88,125.66)(-70.88,-125.66)
        (-43.78, 57.12)(-43.78, -57.12)
    };

    \node[red,font=\small,below] at (axis cs:-9.97, 25.65+1.5) {1};
    \node[red,font=\small,below] at (axis cs: 9.30, 69.81+1.5) {2};
    \node[red,font=\small,below] at (axis cs:-17.09, 40.54+1.5) {3};
    \node[red,font=\small,below] at (axis cs:-29.94, 44.88+1.5) {4};

    \node[black,font=\small,below] at (axis cs:-20.55, 59.84+1.5) {2};
    \node[black,font=\small,below] at (axis cs:-38.76, 39.27+1.5) {3};
    \node[black,font=\small,below] at (axis cs:-70.88,125.66+1.5) {4};
    \node[black,font=\small,below] at (axis cs:-43.78, 57.12+1.5) {1};

    \node[font=\small, black,below] at (axis cs:-40, 135) {STABLE};
    \node[font=\small, black,below] at (axis cs:10, 135) {UNST-};
    \node[font=\small, black,below] at (axis cs:10, 115) {ABLE};

\end{axis}
\end{tikzpicture}

	\else
	\includegraphics[width=\linewidth]{GrowthVsAparGrav.eps}
	\fi
	\caption{The progression of dominant poles in the dynamic system in the s-plane during an Accel-Decel loop for four episodes of key error in the control system: 1. positive jerk, 2. natural frequency transition when accelerating, 3. negative jerk and 4. natural frequency transition when decelerating. The original system (red) and system with implementation of torque feed forward and gain scheduling (black) are shown.}
	\label{fig:Poles plot}
\end{figure}

A second modification, gain scheduling, was introduced to address non-linearities in system stiffness. Moments of high jerk (the derivative of acceleration) require a low proportional gain to move poles leftwards on the s-plane and reduce oscillatory overshoot. Conversely, rejection of disturbances due to variation in rope stiffness is more effective with a high proportional gain. To achieve suitable performance in both scenarios, proportional gains are adjusted both to reduce overshoot following step changes in acceleration and to improve rejection of stiffness disturbances. Figure \ref{fig:AccelData} shows significant improvements in settling time, rise time and disturbance rejection following our redesign of the controller. These improvements are reflected in the leftward shift of the dominant poles in figure \ref{fig:Poles plot}. Under the assumption of a second-order response, each pole was estimated empirically using the damping ratio and natural frequency calculated from acceleration data across episodes of high jerk and natural frequency transition.

\section{Experimental methods} \label{ExpMethod}

\subsection{Experimental setup} \label{ExpSetup}

Our central design case considered that a minimum Atwood number of $A_{\rm t} \approx 0.07$ should achieve a mixed layer width covering the middle half of the tank within the available track length. Sodium chloride salt solution was used as the dense fluid. Propanol-2 was added to fresh water in the less-dense half of the system to match refractive index with that of the salt solution, with $1 \times 10^{-4} \mathrm{\,g\,L^{-1}}$ of Rhodamine 6G fluorescing dye added to enable PLIF measurements. Increasing concentration of propanol-2 increases the refractive index, but does so while decreasing the density of the solution. The refractive indices can therefore be matched without reducing the Atwood number, though because alcohol molecules have a longer carbon chain, mixing may become very weekly non-linear, despite propanol-2 and sodium chloride solutions being fully miscible. When volume fraction gradient vectors are strong and perpendicular to an optical ray, consequent fluctuations in the refractive index can cause ray paths to cross. Refractive indices were measured with a refractometer and the propanol-2 concentration adjusted iteratively. More accurate measurements of fluid density are made using an Anton Paar DMA 4500M density meter and we obtain $A_{\rm t} = 0.0695$, for $14.0 \%$ mass concentration of sodium chloride. As the sodium chloride solution has a higher polarity than the propanol-2 solution, we added dye to the propanol-2 solution. Our approach ensures that the dye remains a reliable proxy for volume fraction estimates throughout the bulk of the flow. Changes in the quantum yield of the fluorescence due to dye aggregation occur on timescales significantly longer than those of the fluid mixing; therefore any systematic error due to dye aggregation is considered local to the fluid interface.

The fluids are stored overnight in sealed and evacuated header tanks to remove dissolved air. The tank is filled using a set of four valved access ports, two of each at the top and bottom of the tank. Fluid is added from the base of the tank and overspill leaves at the top of the tank. Firstly the propanol-2 and dye is added until the tank is full, then the tank is tilted slightly such that any air is removed. This is followed by very slow addition of the salt solution below it, to minimise unwanted pre-mixing, until the two fluids each account for half of the tank volume.

Calibration of the optical diagnostics, discussed in section \ref{PostProcessing}, requires images of the system with the two fluids fully mixed. After each experiment, the tank is emptied, the fluid is manually mixed to ensure full homogeneity and the tank is refilled.

\subsection{Diagnostic post-processing} \label{PostProcessing}

Our PLIF images are of limited use until artefacts of the diagnostic method have been corrected. In pursuit of minimal relative motion between the onboard camera and the tank, the camera is mounted on aluminium struts with the sensor a distance of just $0.2 \mathrm{\,m}$ from the illuminated plane and a fisheye lens is used to capture the full field of view. While helping to prevent vibration, this does however introduce significant lens distortion and vignetting effects, along with refraction effects at the fluid--Perspex and Perspex--air interfaces of the tank. To correct for this, we used a method described in Horne and Lawrie \cite{horne2020aspect} and a flat-field correction technique, respectively. Having considered all the reasons for illumination inconsistencies, there remains a small illumination bias when two lasers overlap which we now correct by normalising.

\begin{figure*}
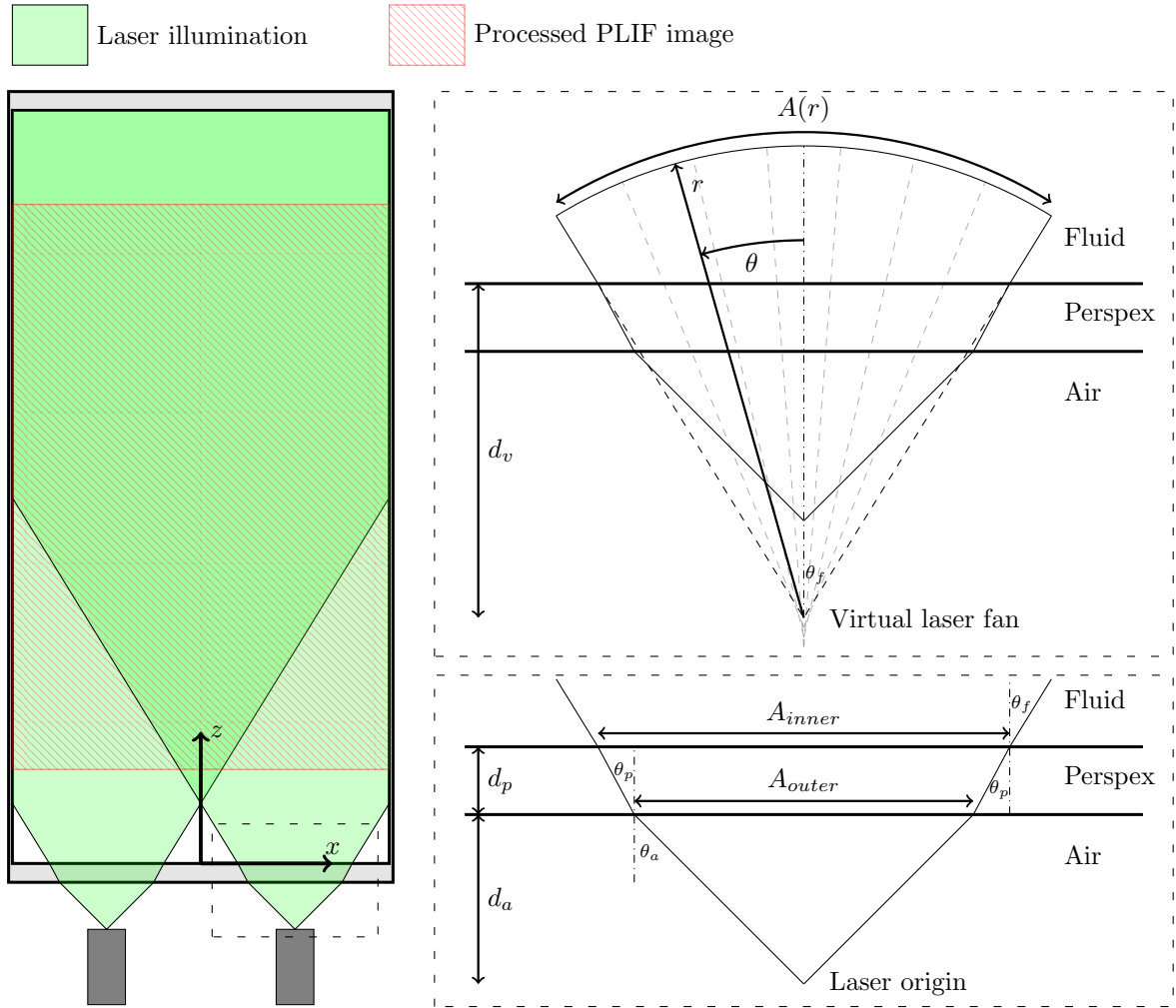

	\centering
	\if\toUseTikz1
	
	\tikzsetnextfilename{LaserIlluminationOld}
	\input{LaserIlluminationOld.tikz}

	\else
	\includegraphics[width=\linewidth]{LaserIlluminationOld.eps}
	\fi
	\caption{The PLIF illumination configuration for the experiments. Planar illumination is provided by a pair of line lasers with $90 \degree$ fans, which are refracted when passing through the Perspex illumination window and into the fluid. The working section of the tank is chosen to give full field illumination.}
	\label{fig:LaserIllumination}
\end{figure*}

A second correction accounts for the variation in intensity of the laser sheet and the response of the dye to that illumination. The fluorescence intensity measured by the CCD, $\iota$, is proportional to the intensity of the light absorbed by the dye, $\Delta I$, according to:
\begin{equation}
\iota = \beta(t,\phi) \Delta I,
\end{equation}
where $\beta$ is the product of geometric collection efficiency, optical system efficiency, and quantum yield, $t$ is the local mixing time and $\phi$ is the volume fraction of the mixed fluids. For reasons discussed in \ref{ExpSetup}, $\beta$ may be considered invariant of $t$ and $\phi$. The laser diodes emit light beams at a constant power, each of which is directed by cylindrical optics to form a fan with angle of $2\theta_a = 90 \degree$ aligned with the tank midplane. The lasers are positioned as shown in figure \ref{fig:LaserIllumination}, offset vertically from the exterior of the tank to optimise coverage of the image plane.
For a single fan laser in free space, the light intensity,$I_{\text{las}}$, at a radial distance $r$ and angle $\theta$ is defined as
\begin{equation} \label{eq:laserIntensity}
I_{\text{las}} \left(r,\theta \right) =
\begin{cases}
\frac{P(\theta)}{2 r \theta_{\rm f} b}, & \mathrm{for} \;\; -\theta_{\rm f} \le \theta \le \theta_{\rm f} \\
0, & \mathrm{otherwise}
\end{cases},
\end{equation}
where $P(\theta)$ is the distribution of laser power incident on a cylindrical surface normal to all light rays and $b$ is plane thickness. The ray fan is bounded within a half-fan angle, $\theta_f$. The illumination intensity therefore reduces \emph{inversely}-linearly with distance from the optics. The intensity is also sensitive to thickness of the laser sheet, which varies between $0.5 - 2.0 \mathrm{\,mm}$ over the height of the tank. However, since dye behaviour is close to linear and the intensity observed by any camera pixel is integrated across small sheet thicknesses, observed intensity may be considered invariant with sheet thickness. Laser light is refracted between the air and the Perspex illumination window and between the Perspex and the fluid.

The tank walls are painted matt black and are assumed to absorb all incident light. The light intensity in free space at any point in the tank is therefore a linear superposition of the direct line of sight intensity contribution from each laser.

\begin{figure}[htbp]
    \centering
    \usetikzlibrary{patterns,angles,quotes,calc}
\begin{tikzpicture}[scale=1.5]
    
    \def\gridwidth{3}
    \def\gridheight{1}
    
    \foreach \x in {0,1,2,3} {
        \draw[thick] (\x,0-\gridheight*0.25) -- (\x,\gridheight*1.25);
    }
    
    \foreach \y in {0,1} {
        \draw[thick] (0-\gridheight*0.25,\y) -- (\gridwidth+\gridheight*0.25,\y);
    }
    
    \node[above,font=\small] at (0.5,\gridheight*1.2) {$j-1$};
    \node[above,font=\small] at (1.5,\gridheight*1.2) {$j$};
    \node[above,font=\small] at (2.5,\gridheight*1.2) {$j+1$};
    
    \node[left] at (0,1.3) {$\iota_{i+1}$};
    \node[left] at (0,0.5) {$\iota_i$};
    \node[left] at (0,-0.3) {$\iota_{i-1}$};
    
    \node[left] at (-0.3,1) {$I_{i+\frac{1}{2}}$};
    \node[left] at (-0.3,0) {$I_{i-\frac{1}{2}}$};
    
    \node[align=center] at (-1.2,0.5) {{Pixel}\\{rows}};
    \node[above] at (1.5,\gridheight+0.5) {{Pixel columns}};
    
    \coordinate (O) at (3.5,-1.2);

    \coordinate (R1end) at (0.6,1.5);
    \coordinate (R2end) at (-0.5,1.5);
    
    \coordinate (Rmidend) at ($(R1end)!0.4!(R2end)$);
    
    \draw[thick] ($(R1end)!+0.15!(O)$) -- ($(O)!+0.35!(R1end)$);
    \draw[thick] ($(R2end)!+0.15!(O)$) -- ($(O)!+0.35!(R2end)$);
    
    \draw[thick, dashed] ($(Rmidend)!+0.15!(O)$) -- ($(O)!0.15!(Rmidend)$);
    
    \coordinate (Vref) at ($($(O)!0.15!(Rmidend)$)+(0,2)$);
    \coordinate (Vstart) at ($(O)!0.15!(Rmidend)$);
    \draw[thick, dashed] (Vref) -- (Vstart);
    
    \pic[draw, angle radius=0.4cm, angle eccentricity=1.5,font=\small, "$\theta_k$"] {angle = Vref--Vstart--Rmidend};
    
    \coordinate (P1) at ($(O)!+0.445!(R2end)$);  
    \coordinate (P2) at ($(R2end)!+0.185!(O)$);  
    \coordinate (P3) at ($(R1end)!+0.185!(O)$);  
    \coordinate (P4) at ($(O)!+0.445!(R1end)$);  
    
    \begin{scope}
        \clip (1,0) rectangle (2,1);
        \fill[pattern=north east lines, pattern color=gray!50] 
            (P1) -- (P2) -- (P3) -- (P4) -- cycle;
    \end{scope}
    
    \draw[thick, black!70!black] (P1) -- (P2) -- (P3) -- (P4) -- cycle;
    
    \node[black!50!black,right,font=\small] at (1.33,0.85) {$f_{l,k}(i,j)$};
    
\end{tikzpicture}
    \caption{Fractional overlap, $f_{l,k}(i,j)$, between ray path $(l,k)$ and pixel grid in cell $(i,j)$ for the staggered grid scheme}
    \label{fig:ray_grid_intersection}
\end{figure}
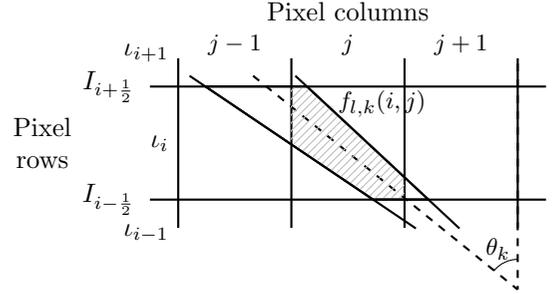

The above is a baseline for variation of laser light intensity in free space, but here we must also account for absorption of light as it passes through the fluid. The Lambert-Beer rule,
\begin{equation}\label{LambertBeer1}
\frac{\partial I}{\partial s} = -\eta ( \hat c)I,
\end{equation}
describes how light intensity, $I$, reduces along a ray path, $s$, according to a dye response function, $\eta(\hat c)$, where $\hat c$ is the concentration normalised by the maximum concentration of dye in the fluid, $\hat c = \frac{c}{c_{\text{max}}}$. If we assume linear dye absorption behaviour with negligible absorption from undyed fluid, $\eta (\hat c) = \epsilon \hat c$, for the dye attenuation constant, $\epsilon$, the variation in light intensity along a ray path is one of exponential decay
\begin{equation}\label{LambertBeer2}
    I^{(i+\frac{1}{2})} = I^{(i-\frac{1}{2})} e^{-\epsilon \int \hat c(s) ds},
\end{equation}
where $I^{(i-\frac{1}{2})}$ and $I^{(i+\frac{1}{2})}$ are the incident intensities at the top and bottom edge of the $i$th pixel row along ray path as shown in figure \ref{fig:ray_grid_intersection}, and $\hat c(s)$ is the variation of dye concentration along the path. 

To combine the effects of laser fan divergence and light attenuation from multiple sources, we implemented a forward ray-based radiative transfer method that exploits fan-beam geometry. The process of density field reconstruction is as follows:

Rays from a source $l$ are emitted and refracted through the air-Perspex and Perspex-fluid boundaries and emerge at angles $\theta_k$ and are evaluated at radial positions $r_i$ to form a ray segment in $(r,\theta)$. These rays intersect the rectilinear pixel grid in $(i,j)$, from which the fractional area of each ray segment that overlaps each pixel is computed from the geometry as $f_{l,k}(i,j)$. The ray intensities $I$ are computed on the staggered grid of pixel boundaries as shown in figure \ref{fig:ray_grid_intersection}. Fractions of ray segments, $f_{l,k}(i,j)$, are used to compute the beam-averaged optical depth, $\tau^{(i)}_{l,k}$ of pixel row $i$,

\begin{equation}
    \tau^{(i)}_{l,k}=\sum_{j} \eta\left(\hat c_{i,j}\right) f_{l,k}(i,j) \Delta s_{l,k}^{(i)},  
\end{equation}
where $\hat c_{i,j}$ is the dye concentration in pixel $i,j$ and $\Delta s_{l,k}^{(i)}$ is the length of the ray segment in pixel units. Within a given pixel, the intensity loss in each ray segment is as follows:
\begin{equation}
    \Delta I_{l,k}^{(i)}=I^{(i-\frac{1}{2})}_{l,k} \left (1-e^{-\tau_{l,k}} \right ).
\end{equation}
This loss $\Delta I_{l,k}^{(i)}$ is redistributed to each pixel $j$ in proportion to each pixel's contribution of the beam-averaged optical depth, $\tau_{l,k}$. The predicted fluorescence intensity, $\iota^{(\text{pred})}_{i,j}$, is therefore given by

\begin{equation}
    \iota^{(\text{pred})}_{i,j}=\sum_{l,k} \left[\Delta I^{(i)}_{l,k} \beta\frac{ \eta(\hat c_{i,j}) f_{l,k}(i,j) \Delta{s^{(i)}_{l,k}}}{\tau_{l,k}} \right]
\end{equation}
 We seek to minimise the error, $\varepsilon$, between the predicted, $\iota^{(\text{pred})}$, and observed, $\iota^{(\text{obs})}$, image intensities.
 \begin{equation}
    \varepsilon_{i,j}= \iota^{(\text{pred})}_{i,j}-\iota^{(\text{obs})}_{i,j},
\end{equation}
Until convergence, a Newton-Raphson update is applied on the concentration field which is of the form
\begin{equation}
    \hat c^{(n+1)}_{i,j}=\hat c^{(n)}_{i,j} - \frac{\varepsilon_{i,j}}{\frac{\partial \varepsilon_{i,j}}{\partial \hat c_{i,j}}}.
\end{equation}
The derivative, $\frac{\partial \varepsilon}{\partial \hat c}$, has two contributions but only $\frac{\partial \iota^{(\text{pred})}}{\partial \hat c}$ is non-zero. We then perform a series expansion in $e^{-\tau_{l,k}}$, and note that for a low-dye concentration $\eta(\hat c)=\epsilon \hat{c}$ where $\epsilon \ll 1$. This implies that the beam-averaged optical depth is thin, we then neglect terms of $\mathcal{O}(\epsilon^{2})$ and higher and obtain
\begin{equation}
    \frac{\partial \varepsilon_{i,j}}{\partial \hat c_{i,j}} \approx \sum_{l,k} \left[I_{l,k}^{(i-\frac{1}{2})}\beta\epsilon f_{l,k}(i,j) \Delta s_{l,k}^{(i)}  \right].
\end{equation}
Rays are sequentially advanced through the pixel domain in $i$ to generate the normalised concentration field. This directly maps onto the volume fraction $\boldsymbol{\phi} \in (0,1)$ from which the density field can be reconstructed.

The intensity $\boldsymbol{I}_{\rm 0}$ of each laser fan is calibrated using two reference images where the density field is known, one taken before and one taken after the experiment. The `before' image corresponds to the initial state of the experiment where the dye occupies only the less dense fluid at the top of the tank. The `after' image is taken after manual mixing to ensure full homogeneity. These images are remapped in $(r,\theta)$ for the relevant source $l$. Using equation \ref{LambertBeer2}, the dye attenuation constant $\epsilon$ is evaluated for the set of `before', $\epsilon_{\rm 1}$, and `after', $\epsilon_{\rm 0.5}$, images.

For each calibration image, the density field is reconstructed using an initial guess of $\rho=0.5$ and refined iteratively for each ray path $k$ until the reconstructed density field converges to the known density field.

For the `before' images, estimation of $\boldsymbol{I}_{\rm 0}$ is limited to ray paths that enter the top half of the tank since there is minimal signal in the region with no dye. To extend the region over which valid intensity distributions can be provided so that `after' images can be used for full-field calibration, $\boldsymbol{I}_{\rm 0}^{(\text{after})}$ is scaled to match the top half of the tank, denoted by the asterisk $(^*)$, with the full field $\boldsymbol{I}_{\rm 0}$ given by
\begin{equation}
        \boldsymbol{I}_{\rm 0}=\boldsymbol{I}_{\rm 0}^{(\text{after})} \frac{\boldsymbol{I}_{\rm 0}^{(\text{before})^*}}{\boldsymbol{I}_{\rm 0}^{(\text{after})^*}}.
\end{equation}
\subsection{Initial condition forcing} \label{InitialConditions}

Acceleration of an initially quiescent stable stratification initialises Rayleigh--Taylor growth with minimal interference at the unstable interface, but the perturbations from which the instability develops are still unknown. In our experiments we seek to better characterise the initial condition by exciting a single mode on the interface before applying destabilising acceleration. Previous experimental work with forced initial conditions have used horizontal \cite{waddell2001experimental,wilkinson2007experimental} or vertical \cite{olson2009experimental,roberts2016effects} oscillation of the fluids to generate a standing wave pattern at the interface. The CAMPI apparatus is designed to apply complex vertical acceleration profiles and is therefore best suited to generation of an initial condition by vertical forcing.

Periodic vertical forcing of a stable fluid stratification can result in a standing wave pattern by excitation of the Faraday instability, first observed by Faraday \cite{faraday1831forms} at a water-air interface. We consider an acceleration on a fluid of form
\begin{equation}
g - f \cos \left(\omega_{\text {drive}} t \right),
\end{equation}
where $\omega_{\text {drive}}$ is angular frequency of oscillations, $g$ is acceleration due to gravity, and $f$ is the magnitude of the forcing. The relationship between wave angular frequency and spatial wavelength can be examined using the dispersion relation. The general form of the dispersion relation for surface gravity waves is
\begin{equation}
\omega^2 = gk \tanh(kh),
\end{equation}
where $k$ is wavenumber, and $h$ is the depth of the fluid. When the waves are at an interface between two fluids of different density in a stable stratification, the dispersion relation uses the \emph{reduced gravity} $\tilde{g} = 2A_{\rm t} g$ \cite{sutherland2010internal}, the interface is assumed to be at the vertical mid-point of a system with total height $2h$, and
\begin{equation}
\omega^2 = \tilde{g}k \tanh(kh).
\end{equation}
The fundamental wavelength for standing waves in a square tank is twice the width of the tank, with the $n^{th}$ harmonic wavelength given by
\begin{equation}
\lambda = \frac{2W}{n},
\end{equation}
where $W$ is the tank width. The height of the tank is twice the width, $h=W$, and the angular frequency of the $n^{th}$ harmonic therefore
\begin{equation}
\omega = \sqrt{\frac{\tilde{g} n \pi \tanh(n \pi)}{W}}.
\end{equation}
A characteristic behaviour of Faraday waves is wave frequency dependent on the forcing frequency. Faraday reported wave frequency at half of the forcing frequency, but more recent work, starting with that of Benjamin and Ursell \cite{benjamin1954stability}, has shown that this is just one of a number of unstable solutions to a Matthieu equation that governs fluid response at an interface.

In this study we have determined forcing frequency required for excitation of a specific harmonic by performing an experimental parameter sweep of frequency.

\begin{figure*}
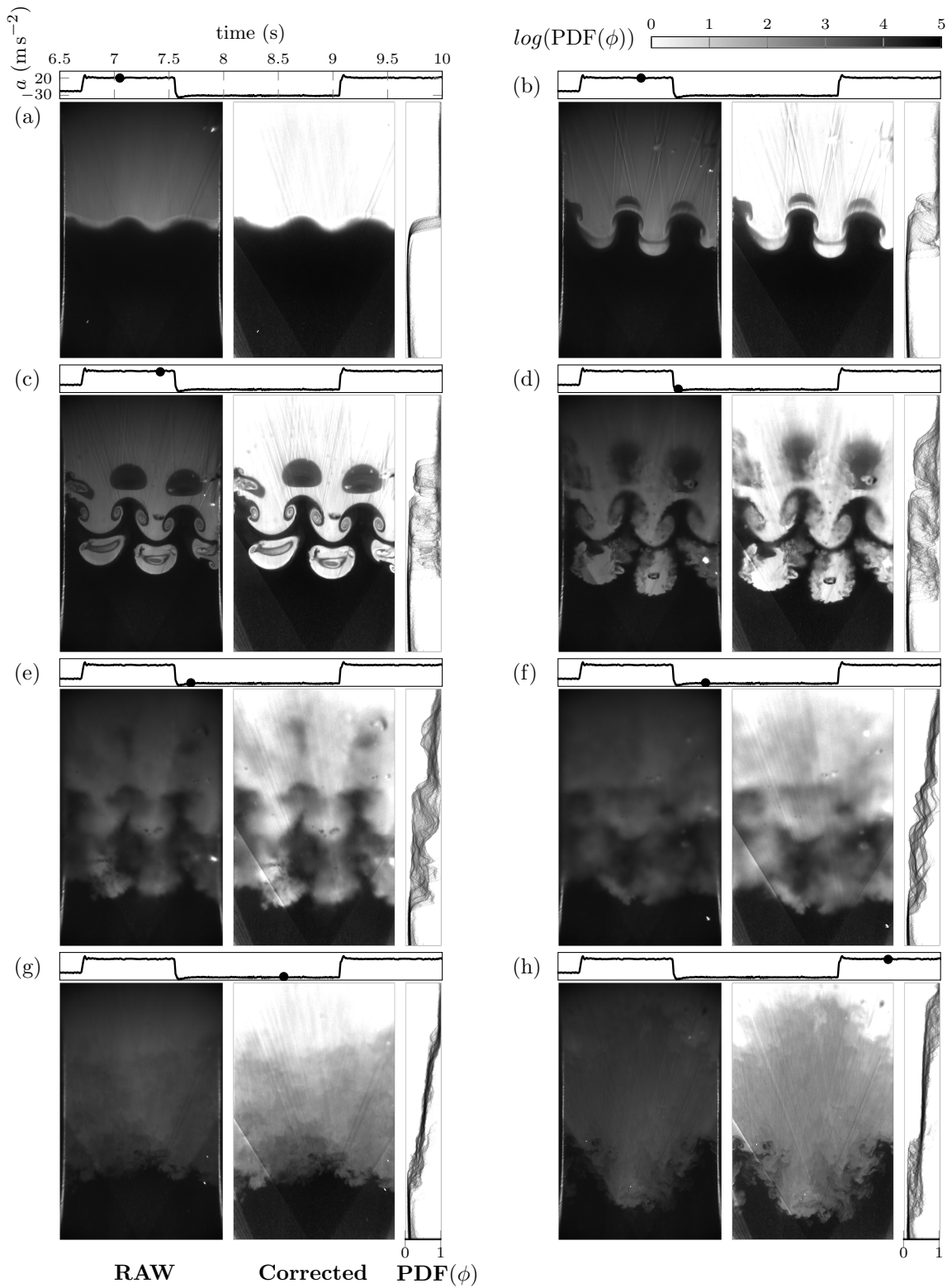

	\centering
	\if\toUseTikz1
	
	\tikzsetnextfilename{ExperimentalResults}
	\input{ExperimentalResults.tikz}

	\else
	\includegraphics[width=\linewidth]{ExperimentalResults.eps}
	\fi
	\caption{Each subfigure shows: the raw experimental image (left), the volume fraction (middle), the probability density function of each z plane sampled in 100 discrete bins (right) and the corresponding acceleration as experienced by the fluid (above, with black dot). Crossing optical paths clearly seen in b) and c) are caused by weekly non-linear mixing locally altering the refractive index.}
	\label{fig:ExperimentalResults}
\end{figure*}

\subsection{Acceleration history} \label{AccelerationHistory}

The acceleration time-history used in our experiment applies vertical oscillations to set an initial condition followed by consecutive periods of acceleration, deceleration, and acceleration. The vertical oscillations have an amplitude of $7.5 {\rm\,m\,s^{-2}}$ $(<9.8 {\rm\,m\,s^{-2}})$ to ensure that the system never becomes Rayleigh--Taylor unstable, and target excitation of the 6\textsuperscript{th} harmonic. This is followed by a period of constant acceleration to accelerate the tank up the track, the sole purpose of which is to allow a longer initial period of Rayleigh--Taylor unstable acceleration. The apparatus targets unstable accelerations with an acceleration as experienced by the fluid of $\hat{g} = 20 {\rm\,m\,s^{-2}}$, and stable decelerations of $\hat{g} = -30 {\rm\,m\,s^{-2}}$. The durations of each period are maximised within the constraints of the track; with initial acceleration over $0.88 \mathrm{\,s}$, deceleration over $1.44 \mathrm{\,s}$, and re-acceleration over $0.88 \mathrm{\,s}$. The apparent gravity measured onboard the tank is plotted in figure \ref{fig:AccelData}, and shows accurate execution of this profile by the apparatus.

\section{Instability behaviour} \label{Results}

The post-processed PLIF diagnostic output provides a measure of volume fraction over a 2D slice at the midplane of the domain. The evolution of volume fraction over the three distinct periods of acceleration is shown in figure \ref{fig:ExperimentalResults}, with images showing a field of view covering the full $0.2 \mathrm{\,m}$ tank width and a height of $0.3 \mathrm{\,m}$ centred at the horizontal midplane, as shown in figure \ref{fig:LaserIllumination}. Images are plotted throughout the Accel-Decel-Accel acceleration history with the first image shortly after the start of the first acceleration episode.

We observe qualitative behaviour consistent with that seen in previous experimental \cite{dimonte2007rayleigh} and numerical \cite{ramaprabhu2013rayleigh,ramaprabhu2016evolution,aslangil2016numerical,aslangil2022rayleigh} studies. The first acceleration episode sees growth of single mode Rayleigh--Taylor instability, with the familiar bubble and spike structure. This structure is quickly lost during deceleration and the mixing region becomes increasingly homogenised. Re-acceleration causes resumption of Rayleigh--Taylor growth, but with a much larger range of length-scales than in the initial acceleration episode. This is due to the complex initial condition with a mixture of density and velocity perturbations across a range of scales caused by the preceding breakdown of coherent structures in the flow.

\begin{figure}
	\centering
	\if\toUseTikz1
	
	\tikzsetnextfilename{MixingGrowth}
	\begin{tikzpicture}

\begin{axis}[
    axis on top=true,
    width=0.49\textwidth,
    xmin=6.5,
    xmax=10,
    ymin=-0.15,
    ymax=0.15,
    ytick={-0.15,-0.1,-0.05,0,0.05,0.1,0.15},
    yticklabels={-0.15,-0.1,-0.05,0,0.05,0.1,0.15},
    xlabel={time (s)},
    ylabel={h (m)},
    tick label style={font=\normalsize},
    label style={font=\large},
    x label style={yshift=-0.0cm},
    x tick label style={yshift=-0.8cm},
]
    \addplot[thick,blue] graphics[xmin=6.5,ymin=-0.15,xmax=10,ymax=0.15] {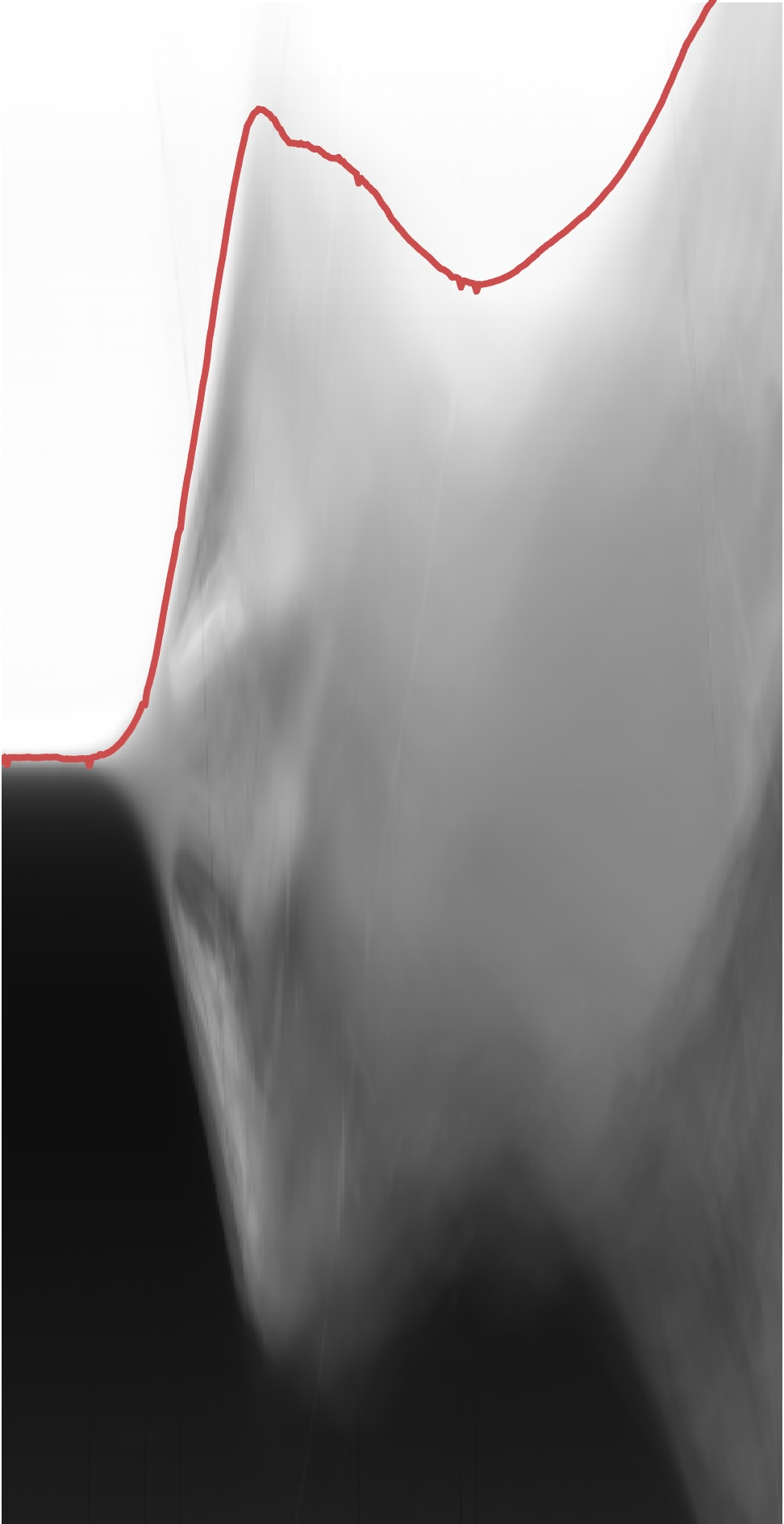};
\end{axis}

\begin{axis}[
    at={(0,0.35\textwidth)},
    axis on top=true,
    width=0.49\textwidth,
    height=0.11\textwidth,
    xmin=0, xmax=1,
    ymin=0, ymax=1,
    ytick=\empty,
    xlabel near ticks,
    xticklabel pos=top,
    xlabel={$\bar{\phi}$},
    tick label style={font=\normalsize},
    label style={font=\large},
]
    \addplot[thick,blue] graphics[xmin=0,ymin=0,xmax=1,ymax=1] {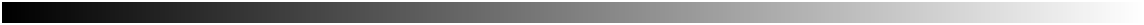};
\end{axis}

\begin{axis}[
    at={(0,-0)},
    anchor=north west,
    yshift=0.0\textwidth-8,
    axis on top=true,
    width=0.49\textwidth,
    height=\textwidth/7.6,
    xmin=6.5, xmax=10,
    ymin=-40, ymax=40,
    xtick={6.5,7,7.5,8,8.5,9,9.5,10},  
    xticklabels=\empty,                 
    ytick={-30,20},
    tick label style={font=\normalsize},
    label style={font=\large},
    ylabel={$a {\,\rm\,(m\,s^{-2}})$)}
]
    \addplot graphics[xmin=6.5,ymin=-40,xmax=10,ymax=40] {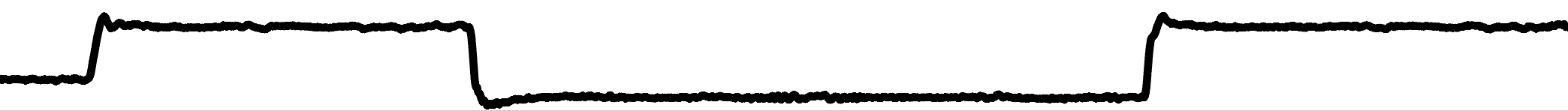};
\end{axis}

\end{tikzpicture}

	\else
	\includegraphics[width=\linewidth]{MixingGrowth.eps}
	\fi
	\caption{The evolution of horizontally averaged volume fraction. An integral measure of mixing region width, $h(t) = 6 \int \bar{\phi} (1 - \bar{\phi}) dz - h_0$, is superimposed in red where $h$ is the height above the interface and $a$ is the corresponding acceleration history.}
	\label{fig:MixingGrowth}
\end{figure}

The growth of the mixing region is studied quantitatively using two measures, both of which are plotted in figure \ref{fig:MixingGrowth}. The background image shows evolution of horizontally averaged volume fraction, $\bar{\phi}(z,t)$, with time axis consistent with the acceleration data in figure \ref{fig:AccelData}. The superimposed red line plots $h(t) = 6 \int \bar{\phi} (1 - \bar{\phi}) dz - h_0$, the measure of Andrews and Spalding \cite{andrews1990RTexperimental2D} with a standard scaling and a constant offset to negate contributions from diagnostic noise as measured at $t=0$.

In the first acceleration episode we observe the mixing region growth expected of single mode Rayleigh--Taylor, with initial exponential growth reaching a wavenumber specific terminal velocity. Then with the transition to the deceleration phase the direction of the buoyancy forces reverses, accelerating the bubbles and spikes back towards the centre of the tank. These structures collide and break down, forming the more homogeneous mixing region visible between $t=7.6 \mathrm{\,s}$ and $t=8.5 \mathrm{\, s}$ in the plot in figure \ref{fig:MixingGrowth}. The integral measure demonstrates shrinking of the mixing region due to the loss of prominent structures of unmixed fluid, the first demonstration of such behaviour in experiments.

\section{Conclusions} \label{Conclusions}

The sensitivity of Rayleigh--Taylor instability to varying initial conditions and acceleration histories is an important open question in industrial applications and natural scenarios, and useful analysis of this behaviour must be derived from a known ground truth. We have therefore performed an \emph{experimental} study of Rayleigh--Taylor instability driven by consecutive periods of Rayleigh--Taylor unstable, stable, and unstable acceleration, the first such experimental study on a fully miscible, low Atwood number fluid system.

To perform this study and to answer a long queue of further research questions, we have developed the CAMPI apparatus, which facilitates well-diagnosed experiments of Rayleigh--Taylor instability subject to a wide range of prescribed initial conditions and acceleration histories. We have presented the design of the apparatus, including the highly controllable two-directional winching assembly and the high resolution moving reference frame diagnostic system. We have also presented the experimental methods and post-processing techniques that allow good understanding of the phenomena influencing the experiment and generate time evolving field data of the output quantities of interest.

An initial experiment has been conducted investigating the behaviour of the instability through an Accel-Decel-Accel acceleration history. This has replicated behaviour seen in previous numerical studies, most notably with shrinking of the mixing region width throughout the episode of deceleration. The de-coherence of previously coherent structures and homogenisation of the mixing region are qualitatively evident in raw volume fraction diagnostic images and are confirmed quantitatively by measures of mixing region evolution computed in post-processing. We present these results to demonstrate the unique capability that the CAMPI apparatus can provide to the field, as a source of ground truth data on the behaviour of Rayleigh--Taylor instability across a wide range of physical regimes.

\bmhead{Acknowledgments}
We would like to thank AWE for their support and encouragement. We would also like to thank the staff at the University of Bristol whose work has made the success of the CAMPI apparatus possible, most notably Tim Bond and Andy Brown.

\bmhead{Author contributions}
Conceptualisation: all authors; Formal analysis: all authors; Methodology: all authors; Software: J. T. Horne-Jones and D. J. Glinnan; Supervision: A. G. W. Lawrie and R. J. R. Williams; Visualisation: J. T. Horne-Jones and D. J. Glinnan; Writing - original draft: all authors; Writing - review \& editing: all authors.

\bmhead{Funding}
This work was funded under grant: EPSRC14220081/EP/M507337/1/AWE30331919/0.

\bmhead{Data availability}
All available data is contained within the paper.

\noindent \textbf{UK Ministry of Defence \textcopyright{}  Crown Owned Copyright 2025/AWE}

\section*{Declarations}

\bmhead{Conflict of interest} The authors declare they have no conflict of interest.


\bibliography{References}


\end{document}